\documentclass[12pt]{article}

\usepackage{axodraw}

\def\dslash{\raisebox{1pt}{$\slash$} \hspace{-7pt} \partial}
\def\Dslash{\raisebox{1pt}{$\slash$} \hspace{-8pt} D}
\def\bea{\begin{eqnarray}}
\def\eea{\end{eqnarray}}
\def\be{\begin{equation}}
\def\ee{\end{equation}}
\def\nn{\nonumber}
\def\a{& \hspace{-7pt}}

\begin{document}

\thispagestyle{empty}

\begin{center}
\hfill CERN-TH/2001-193\\
\hfill ROMA-1319/01 \\
\hfill SISSA-58/2001/EP \\

\begin{center}

\vspace{1.7cm}

{\LARGE\bf Non-local symmetry breaking \\

\vskip 5pt 

in Kaluza-Klein theories}

\end{center}

\vspace{1.4cm}

{\sc A. Masiero$^{a}$, C.A. Scrucca$^{b}$, M. Serone$^{a}$ and
L. Silvestrini$^{c}$} \\

\vspace{1.2cm}

${}^a$
{\em ISAS-SISSA, Via Beirut 2-4, I-34013 Trieste, Italy} \\
{\em INFN, sez. di Trieste, Italy} \\
{\footnotesize \tt masiero@sissa.it, serone@sissa.it}
\vspace{.3cm}

${}^b$
{\em CERN, 1211 Geneva 23, Switzerland} \\
{\footnotesize \tt Claudio.Scrucca@cern.ch}
\vspace{.3cm}

${}^c$
{\em INFN, sez. di Roma, Dip. di Fisica, Univ. di Roma ``La Sapienza'',} \\
{\em P.le Aldo Moro 2, I-00185, Roma, Italy}\\
{\footnotesize \tt Luca.Silvestrini@roma1.infn.it }
\end{center}

\vspace{0.8cm}

\centerline{\bf Abstract}
\vspace{2 mm}
\begin{quote}\small
Scherk-Schwarz gauge symmetry breaking of a $D$-dimensional field theory 
model compactified on a circle is analyzed. It is explicitly shown that 
forbidden couplings in the unbroken theory appear in the one-loop
effective action only in a non-local way, implying that they are
finite at all orders in perturbation theory. This result can be 
understood as a consequence of the local gauge symmetry, but it holds 
true also in the global limit. 

\end{quote}

\vfill

\newpage
\setcounter{equation}{0}

There has been recently a vast amount of work exploring the
implications of breaking gauge \cite{gauge} or supersymmetries
\cite{Antoniadis:1990ew,antonia,barbieri} via the Scherk-Schwarz (SS)
mechanism \cite{ss}.  Indeed, this mechanism, which until recently had
not been put into action in explicit theories, has revealed exciting,
but also puzzling aspects.  In particular, amazing ultraviolet
softness properties in theories with extra dimensions where
supersymmetry (SUSY) breaking occurs through the SS mechanism have
sparkled a vast debate on what is at the root of such ``finiteness
protection'' (a protection which is even more efficient than that
provided by softly broken supersymmetry in ordinary four-dimensional
theories).

The physical observables which have first drawn the attention on this
apparently surprising ultraviolet behaviour of extra-dimensional
supersymmetric theories are the scalar masses \cite{antonia}.
Although the higher-dimensional theory is generally
non-renormalizable, it was pointed out that the radiative corrections
to the scalar masses are quite insensitive to the cutoff $\Lambda$ if
the compact space is of size $O({\rm TeV}^{-1})$. However the
rationale for such finiteness is still debatable \cite{debate}. At a
more technical level, doubts were cast on the physical meaning of the
commonly adopted ``Kaluza-Klein (KK) regularization'' consisting in
summing over the infinite tower of KK states and performing the
one-loop integral over the momenta. It was argued that the quadratic
dependence of the Higgs masses on the ultraviolet cut-off could
reappear in higher orders in perturbation theory given the {\it ad
hoc} character of the ``KK regularization''.

This Letter aims at clarifying the debate by analyzing the SS breaking
mechanism in a simpler gauge model in $D$ dimensions compactified on a
circle down to $D-1$ dimensions with arbitrary boundary conditions.
We compute one-loop corrections to scalar masses and explicitly show
that the SS breaking manifests itself only via the appearance of terms
which are {\em non-local} along the compact dimension.  Although we
cannot provide an equivalent description for orbifolds and for 
the case in which the SS mechanism, together with projections, 
is invoked to break SUSY, we
think that the main idea is basically the same, namely the Higgs mass
correction is finite because it is necessarily a non-local term in the
one-loop effective action \footnote{The non-locality of the SS breaking 
mechanism was already pointed out in \cite{ahnsw}, but its implications 
for the effective Lagrangian were not fully exploited.}.

Here are our main results. First we observe that to get a sensible
result the sum over the whole tower of KK states has to be performed,
since any acceptable regularization has to respect the local
symmetries in $D$ dimensions which are not actually broken when the SS
mechanism takes place.  The Ward identities can be recovered as in the
unbroken case, but with twisted gauge parameters. Alternatively,
following \cite{hos}, the SS mechanism can be viewed as a spontaneous
breaking through the vacuum expectation value (VEV) of $A_D$, namely
the component of the connection along the compact coordinate. Any
truncation of the sum would correspond to a regularization which does
not respect the symmetries of the theory. Second and more important of
our results is that the non-local nature of the SS breaking appears in
the Lagrangian through non-local terms. More precisely, terms that
would be forbidden in the unbroken theory may now appear, but only as
non-local ones, associated with Wilson lines and Wilson loops.  
Hence, for such terms,
the breaking \`a la SS corresponds to a non-local spontaneous symmetry
breaking. The non-local symmetric terms in the $D$-dimensional
Lagrangian correspond to the non-symmetric local terms of the
spontaneously broken theory in $D-1$ dimensions. As an illustration of
this result, we analyse the mass splitting in a $SU(2)\times
U(1)$ scalar doublet and its generalization for a $SU(N)$ model.

The non-locality of these gauge invariant terms arising from the SS
breaking is at the root of the finiteness of the radiative corrections
we compute.  
In a renormalized quantum field theory, at any order in
perturbation theory, all divergent terms are polynomial in the
external momenta. Therefore, they correspond to local counterterms to
be added to the lagrangian. Since the only symmetry breaking
quantities generated by radiative corrections in this model correspond
to non-local gauge invariant terms, they cannot be divergent. This
argument can be also applied to non-renormalizable field theories.
Hence we can conclude that the finiteness  of radiatively induced
symmetry breaking terms has to persist
at all orders in perturbation theory. An important comment is however
in order. Our results show that any symmetry-breaking term, being
necessarily non-local, is finite, but this does not imply at all that
the physical values of the associated quantities are finite and
cut-off independent. In the case we study, of course, divergences
appear in local gauge-invariant terms.  Being that the underlying theory
is non-renormalizable (for $D> 4$), a certain sensitivity on the cut-off
is therefore present and unavoidable,
but it is confined to local gauge-invariant quantities.

Let us then consider the following $D$-dimensional $SU(2)\times U(1)$
gauge theory (the reason of the additional $U(1)$ factor will be clear
later on):
\be
{\cal L}=(D_M \Phi)^\dagger (D^M \Phi) + \bar \Psi i \Dslash \Psi 
+ \bar \chi i \dslash \chi - \frac{1}{8g^2}\, \mathrm{Tr}\, F_{MN}^2
-\frac{1}{4g^{\prime 2}}\, F_{MN}^{\prime 2}
+  \lambda ( \Phi^\dagger \bar \chi \Psi + \mbox{h.c.}), 
\label{eq:lagrangian}  
\ee
where $\Phi$ and $\Psi$ are $U(1)$-charged $SU(2)$ doublets and $\chi$
is a singlet, $D_M=\partial_M - i A^a_M T^a -i A^0$, with $T^a$ the
Pauli matrices and $A^0$ the $U(1)$ connection, $M=(\mu,D)$,
$\mu=0,\dots,D-2$ and $\Dslash= D_M \gamma^M$.

We take the $D^{{\rm th}}$ dimension to be a circle of radius $R$. The
SS breaking is achieved by imposing the following periodicities on the
fields in eq.~(\ref{eq:lagrangian}):
\begin{equation}
\begin{array}{cc}
\Phi(y + 2 \pi R)= U\,\Phi(y) & 
\Psi(y+2 \pi R)=U\,\Psi(y) \\
A_M(y+2 \pi R) = U\,A_M(y)\, U^\dagger &
\chi(y+2 \pi R)=\chi(y)\,,
\end{array}
\label{eq:period}
\end{equation}
where $y$ denotes the compact coordinate and
\begin{equation}
U\equiv U(\alpha, \beta)=e^{2 i \pi ( \alpha T_3 + \beta ) }   \,.
\label{U}
\end{equation}
The corresponding Fourier decomposition of a generic field $\phi$ along the
compact dimension is given by:
\begin{equation}
\phi(x,y)=\frac{1}{\sqrt{2 \pi R}} \sum_{n=-\infty}^{+\infty} e^{i(n+a)y/R}
\phi_n(x)\,,
\end{equation}
where $x$ represents the $(D\!-\!1)$-dimensional coordinates. 
The values of $a$ for
the above fields are as follows: $A^{0,3}$ and $\chi$ remain periodic ($a=0$),
$a=\alpha+ \beta$ for the upper components $\phi_1$ and $\psi_1$ 
of the doublets, $a=-\alpha+ \beta$ for the lower components $\phi_2$ 
and $\psi_2$ and $a=\pm 2 \alpha$
for $A^\pm\equiv (A^1 \pm i A^2)/\sqrt{2}$.
These periodicities are
conserved by any single-valued gauge transformation on the circle. 

It is interesting to note that if the gauge parameters
$\varepsilon_a(x,y)$ have the same periodicity as the corresponding gauge 
fields in eq.~(\ref{eq:period}), the compactified theory is still invariant
under the {\em full} $SU(2)\times U(1)$ gauge group. Of course, since 
$\varepsilon_{1,2}(x,y)$ do not admit the rigid limit in $y$, from the 
$(D\!-\!1)$-dimensional point of view the unbroken gauge group is only
$U(1)\times U(1)$.
Furthermore, as shown by Hosotani \cite{hos}, one can perform more general 
gauge transformations, changing the boundary conditions for the fields in
eq.~(\ref{eq:period}). In particular, it is possible to obtain the case 
in which all fields are periodic, by means of the following gauge 
transformation:
\begin{equation}
\Omega(y)=e^{-i (\alpha \,T_3 + \beta) y/R }  \,.
\end{equation}
However, the $D^{\rm th}$ component of the transformed gauge fields 
acquires a vacuum expectation value given by 
\be
\langle A_D \rangle=- (\alpha \, T_3 + \beta )/R \,.
\label{advev}
\ee
Indeed, one can check that this VEV reproduces the mass spectrum 
obtained from eq.~(\ref{eq:period}).
Therefore, the SS breaking of a gauge symmetry can be alternatively 
seen, when all fields are periodic, as a spontaneous symmetry
breaking induced by the VEV of $A_D$.

Since the gauge symmetry is broken in a peculiar way through the boundary 
conditions or, alternatively, through a VEV for $A_D$ \footnote{Notice that 
also this last picture makes sense due to the presence of a compactified 
coordinate. In absence of the latter, constant gauge connections can always 
be gauged away.}, we expect that $SU(2)$-breaking terms should be non-local 
around the compact coordinate. 
It is our aim to explicitly show that this is indeed what happens, by 
considering mass terms induced in the one-loop effective action 
for the $\Phi$ doublet.
We compute the two-point functions for on-shell external momenta, 
$q^2=q_D^2$, in order to extract directly
the renormalized mass terms. 

To better discuss the global limit $g,g^\prime\rightarrow 0$ of 
our results, we will first compute the fermionic contribution to the 
two-point function, that we denote by ${\cal F}_a(q^2,q_D)$,
where $a=\beta \pm \alpha$ for the doublet. 
This is given by
\be
\label{Aglobale}
i\, {\cal F}_a(q^2,q_D) 
= \frac{\lambda^2 2^{[D/2]}}{2\pi R} \int\! \frac{d^{D-1}p}{(2\pi )^{D-1}} 
\sum_{p_D} \frac{p\cdot (p+q) - p_D (p_D+q_D)}
{(p^2-p_D^2)[(p+q)^2-(p_D+q_D)^2]} 
\,, 
\ee
where 
$[D/2]={\rm int (D/2)}$ takes into account the dimensionality of the 
gamma matrices in $D$ dimensions. 
The sum over the KK tower can be easily performed 
and one finds, going to the Euclidean, at $q^2=q_D^2$:  
\be
{\cal F}_a(q_D^2,q_D) =
- \frac {\lambda^2 2^{[D/2]-1}R^{2-D}}{(4\pi)^{\frac {D-1}2} 
\Gamma(\frac {D-1}2)}\,
\mbox{Re}\, \int_0^\infty \!\! 
dx \, {x^{D-3}} \Big[\coth \pi x + \coth \pi (x + i\,a) \Big] \,,
\label{globale1}
\ee
where $x= pR$.
Interestingly, the result does not depend on the KK level $q_D=(m+a)/R$
and, hence, each KK mode $\Phi_m$ of the doublet gets a universal 
one-loop correction to its mass. 
This property is a consequence of the $D$-dimensional local gauge 
invariance that is still present in the theory, and would have been rather 
obscure from a purely $(D\!-\!1)$-dimensional point of view.

It is convenient to define 
${\cal F}_{\pm} = {\cal F}_{\beta+\alpha} (q_D^2,q_D) 
\pm  {\cal F}_{\beta-\alpha}(q_D^2,q_D)$, encoding respectively 
the $SU(2)$-symmetric and $SU(2)$-breaking parts. 
The momentum integral can be explicitly evaluated and one obtains:
\bea
{\cal F}_{+} \a=\a 
\lambda^2 2^{[D/2]+1}
\left[\Lambda^{D-2} + \sum_{n=1}^\infty C_D^{(n)}
{\cos 2\pi n\alpha \,\cos 2\pi n \beta} \right] \,,\nn \\
{\cal F}_{-} \a=\a - \lambda^2 2^{[D/2]+1}  
\sum_{n=1}^\infty C_D^{(n)} \sin 2\pi n\alpha \,\sin 2\pi n \beta \,,
\label{Apm}
\eea
where $C_D^{(n)}=(D-3)!/[(4\pi)^{\frac {D-1}2} \Gamma(\frac {D-1}2)
(2\pi Rn)^{D-2}]$. The $(\alpha,\beta)$-independent term in (\ref{Apm}) 
represents the usual divergence that one gets in the unbroken case. 
Since gauge invariance does not forbid the appearance of such term, 
it is generated with the expected degree of divergence. 

The form of the one-loop induced 
$SU(2)$-breaking mass term for $\Phi$ can now be derived by taking
the Fourier transform of (\ref{Apm}). It yields 
the following non-local coupling 
\begin{equation}
{\cal L}_{{\rm nl}}^{\cal F}= \lambda^2 
2^{[D/2]-1} \sum_{n=1}^{\infty} 
C_D^{(n)} \,\Phi^\dagger (y)
\, W_n \,
\Phi(y+2\pi R n) + \mbox{h.c.} \,,
\label{nonlocalmass}
\end{equation}
involving Wilson lines around the compact direction:
\begin{equation}
W_n=W_n(y)= {\cal P} 
e^{i \int_y^{y+2\pi R n} \!\!  \ A_D(y^\prime)\,{\rm d}y^\prime}.
\label{eq:wline}
\end{equation} 
In the picture in which $\langle A_D \rangle = -(\alpha T_3 + \beta)/R$ 
and the boundary conditions are periodic, one finds
\begin{equation}
{\rm Re}\, \langle W_n \rangle  = 
\cos 2\pi n\alpha \cos 2\pi n \beta - 
T_3 \sin 2\pi n\alpha \sin 2\pi n \beta \,,
\nn
\end{equation}
reproducing the correct $(\alpha, \beta)$ dependence found in (\ref{Apm}). 
Alternatively, in the picture in which $\langle A_D \rangle = 0$ but the 
boundary conditions are twisted, one has 
\begin{equation}
\Phi (y+2\pi R n) = U^n(\alpha, \beta) \Phi (y) \,,
\label{Phiper}
\end{equation}
and the same result is obtained, as expected.  Eq.~(\ref{nonlocalmass}) 
shows explicitly that $SU(2)$-breaking terms must
be non-local around the compact direction, and hence finite.

Let us now consider the gauge contribution to the two-point function, denoted
by ${\cal G}_a(q^2,q_D)$. 
The general formula can be written as
\be
i\,{\cal G}_a(q^2,q_D) 
=\frac{1}{2\pi R} \int\! \frac{d^{D-1}p}{(2\pi )^{D-1}} 
\sum_{colors} \sum_{p_D} \Bigg[\frac{-2D}{p^2-p_D^2} +
\frac{(p+2q)^2 - (p_D+2q_D)^2}
{(p^2-p_D^2)[(p+q)^2-(p_D+q_D)^2]} \Bigg] , 
\label{Alocale}
\ee
omitting the group generators and couplings and the $p_D$
dependence on the colors.
Defining symmetric and breaking parts ${\cal G}_{\pm}$ as
before and fixing $q^2=q_D^2$, one finds for $SU(2)\times U(1)$: 
\begin{eqnarray}
{\cal G}_{+} &=& -4
\sum_{n=1}^\infty C_D^{(n)} \, \biggl[ (3 g^2 + g^{\prime 2})
\cos 2\pi n\alpha \,\cos 2\pi n \beta + 
4 g^2 (D-1) \cos 4 \pi n \alpha \biggr]\,, \nn \\
{\cal G}_{-} &=& 4 (g^{\prime 2} - g^2) 
\sum_{n=1}^\infty C_D^{(n)} \sin 2\pi n\alpha \,\sin 2\pi n \beta \,,
\label{A-G}
\end{eqnarray}
where we omitted a $(\alpha,\beta)$ independent divergence in ${\cal G}_+$.
Notice that ${\cal G}_{\pm}$ do not have precisely to coincide with those
of eq.~(\ref{Apm}), because they mix up with Wilson loop contributions.
In fact, these represent clearly another class of terms that are
finite because non-local around the compact dimension (see also
\cite{hcg}). New non-local structures are therefore present.  In
particular, the effective coupling gets also Wilson loop
contributions:
\begin{eqnarray}
{\cal L}_{\rm nl}^{\cal G} &=& 
-\sum_{n=1}^{\infty} C_D^{(n)} \Bigg[
\Big(g^{\prime 2} - g^2 + 4 \, g^2 (D-1)\,{\rm Tr}\,W_n^\dagger \Big) 
\, \Phi^\dagger (y)\, W_n \, \Phi(y+2\pi R n) \nn \\
&\;&\hspace{57pt} + \, 2 \, g^2 {\rm Tr}\,W_n \Phi^\dagger (y) \Phi(y) \Bigg]
+ {\rm h.c.} \,. 
\label{eq:nonlocalgauge}
\end{eqnarray}
Notice that for $g=g^\prime$ ($U(2)$ case), ${\cal G}_-$ vanishes and
for the pure $SU(2)$ case, i.e. $\beta=0$, both ${\cal G}_-$ and
${\cal F}_{-}$ vanish (this is a general feature of $SU(2M)$ groups; 
see below). No $SU(2)$ breaking terms are generated in this case.  

Although our analysis was restricted to the one-loop approximation, it 
is clear that the non-locality of symmetry-breaking couplings will 
persist at any order of perturbation theory, never allowing for the 
appearance of UV divergences. These occur in configuration space at 
very short distances, but all the effective symmetry-breaking interactions 
involve Wilson lines winding around the compact direction and can never 
produce short-distance singularities. Indeed, we argued that radiative
corrections preserve the symmetries of the $D$-dimensional theory, and
that the Wilson line is the only quantity that can be generated
leading to symmetry-breaking terms from the $(D-1)$-dimensional point
of view. Thus, any higher order non-symmetric contribution will be a
function of the Wilson line, and it will be finite once all the
relevant symmetric counterterms of lower order are added to all
subgraphs.  

Let us now consider the global limit $g,g^\prime\rightarrow 0$.
In this case, the gauge field is clearly non-dynamical but its
VEV (\ref{advev}), which is $g,g^\prime$-independent and still induces
a non-trivial value of $W_n$,
can be considered as a sort of left-over flux or condensate, responsible 
for the twist in the boundary conditions of the fields.
The gauge contribution (\ref{eq:nonlocalgauge}) now clearly vanishes and
eq.~(\ref{nonlocalmass}) reduces to
\begin{equation}
{\cal L}_{{\rm nl}}= \lambda^2 2^{[D/2]-1} \sum_{n=1}^{\infty} 
C_D^{(n)} \Phi^\dagger (y)
\Phi(y+2\pi R n) + \mbox{h.c.} \,,
\label{nonlocalmassG}
\end{equation}
with the periodicities as in (\ref{Phiper}).  It is interesting to
notice that the terms in (\ref{nonlocalmassG}) can be alternatively
considered as non-local and (global) $SU(2)$-invariant, written as
above, or local and $SU(2)$ breaking using (\ref{Phiper}). It is
clear, however, that finiteness is also in this case ensured at all
orders in perturbation theory.

An alternative approach to the one followed here consists in computing
the two-point functions at $q=0$ and generic $q_D=(m+a)/R$. In this
way, one gets an explicit $(m+a)$-dependence in the amplitudes that,
when the latter is reinterpreted as $-i\partial_D$, leads to an
expansion in derivatives for an infinite number of terms. We checked
that again all the $SU(2)$ breaking terms are finite. Moreover, as
expected, the leading $q_D=0$ mass term coincides with (\ref{Apm}) and
(\ref{A-G}).  

The analysis that we have performed for the $SU(2) \times U(1)$ model can be
easily extended to a general $SU(N)$ model with (\ref{U}) given by 
$U=\exp (2i \pi \alpha T)$, where $T={\rm diag}(t_1,\dots,t_{N})$ and 
$t_{1,\dots,b}=a/c$, $t_{(b+1),\dots,N}=-b/c$, $c=\sqrt{(b a^2+ a b^2)/2}$. 
The fermionic contribution eq.~(\ref{globale1}) and the discussion following 
it remain valid, with $a$ in eq.~(\ref{globale1}) being now the twist in 
$q_D$ of each component of the $N$-plet. For the gauge contribution,
one finds instead the following effective coupling for the $\alpha$-dependent
part:
\begin{eqnarray}
{\cal L}_{\rm nl}^{SU(N)} &=& 
-g^2 \sum_{n=1}^{\infty} C_D^{(n)} \Bigg[
\Big(\!-\!\frac{2}{N}+ 4(D-1)\,{\rm Tr}\,W_n^\dagger\Big)  
\Phi^\dagger (y)\, W_n \, \Phi(y+2\pi R n) \nn \\
&\;&\hspace{65pt} + \, 2\, {\rm Tr}\,W_n \Phi^\dagger (y) \Phi(y) \Bigg]
+ {\rm h.c.} \,. 
\end{eqnarray}
For $N=2M$, one has $a=b=M$ so that ${\rm Re}\,W_n$ is an even
function of $\alpha$ and ${\rm Tr}\,W_n$ is real. It follows that no mass 
splitting between the upper and lower components of the $2M$-plet is 
radiatively generated. On the contrary, for $N=2M+1$, one has $a=M$, $b=M+1$ 
and a mass splitting between the upper $M+1$ and lower $M$ components is 
produced at one-loop. 

Similarly, one can also study the one-loop amplitude $\langle
A_D\,A_D\rangle$. For example, in the $SU(2) \times U(1)$ model, 
one finds that all KK levels but the zero modes of
$A_D^0$ and $A_D^3$ do not receive any correction, whereas the latter
get a finite and $(\alpha,\beta)$-dependent correction. This is what
is expected. Gauge symmetry forbids the appearance of a mass term for
$A_M$, whereas a one-loop effective potential for $A_{D,0}^{0,3}$ is
generated. These zero modes are the non-integrable phases of
\cite{hos}, related to the eigenvalues of the Wilson loop.

Our results hold by summing over the entire tower of KK states.
One can easily verify that new and $(\alpha, \beta)$ dependent 
divergences appear (for $D > 4$), also in symmetry-breaking terms,  
as soon as the sum over KK modes is truncated. This is due to
the fact that the local gauge symmetry along the 
compact coordinate is explicitly broken in this way. 

As we already mentioned, we did not consider the case in which the SS
mechanism is implemented through R-symmetry transformations to break
SUSY and induce a finite Higgs mass \cite{barbieri}.  An explicit
computation in this context, along the lines considered here, would be
very interesting.  An important and related open issue is to
understand whether the SS mechanism applied to an R-symmetry can be
viewed as a spontaneous breaking of local supersymmetry, very much
along the lines of \cite{hos}.  Recent work suggests that this might
actually be the case \cite{Marti}, but a full supergravity analysis is
required to definitively establish this result.

\section*{Acknowledgments}

We would like to thank N. Arkani-Hamed, R. Barbieri, A. Donini,
P. Gambino, H. Gies, L. Griguolo, G. Isidori and F. Zwirner for 
interesting discussions. This work has been partially supported by the 
EEC through the RTN network ``Across the Energy Frontier'', contract 
HPRN-CT-2000-00122.

\end{document}